# "No High Energy Emission" GRB Class Is Attributable to Brightness Bias

J. T. Bonnell [1,2] and J. P. Norris [1]

[1] NASA/Goddard Space Flight Center, Greenbelt, MD 20771.
[2] Universities Space Research Association.

## Abstract

The inhomogeneous brightness distribution of BATSE detected gamma-ray bursts has been considered strong evidence for their cosmological origin. However, subclasses of gamma-ray bursts have been shown to have significantly more homogeneous brightness distributions. Pendleton et al. (1997) have found such a result for gamma-ray bursts with no detectable emission at energies >300 keV. Accordingly, it has been suggested that these no high energy (NHE) emission bursts represent an underluminous population of nearby sources. A distinct homogeneous NHE brightness distribution has also been considered as evidence for beaming of different spectral components of the prompt burst emission. We synthesize observed distributions of gamma-ray bursts based on a sample of typical bright BATSE bursts with intrinsic high energy emission and adopt a single cosmological distance scale for all sources. We find that the resulting synthetic NHE bursts do indeed have a more nearly homogeneous intensity distribution when an appropriate decrease in signal to noise and redshifted spectrum is incorporated. We argue that the definition of NHE bursts, and soft-spectrum bursts in general, naturally produces a steep distribution. The NHE class of gamma-ray bursts is therefore likely due to brightness bias.

# 1. INTRODUCTION

The inhomogeneous brightness distribution of gamma-ray bursts (GRBs) coupled with their isotropic sky distribution as observed by BATSE (Meegan et al. 1992) were an early indication that GRB sources lie at cosmological distances. Recently, measured redshifts for GRB optical counterparts (e.g. Metzgar et al. 1997; Djorgovski et al. 1998; Bloom et al. 1998) have produced dramatic confirmation of the cosmological distance scale for a representative set of GRBs.

Each successive BATSE GRB catalog (e.g. Paciesas et al. 1999) has revealed a slope for the cumulative brightness distribution, log N – log PF, which falls significantly below a –3/2 power law at low peak fluxes, where N is the number brighter than a given peak flux (PF). The -3/2 law would be expected if the GRB rate were globally constant and the sources strewn homogeneously in a Euclidean space. Generally the observed departure from a -3/2 slope at low peak fluxes should be a natural consequence of a population of sources extending to cosmological distances, thus sampling curved spacetime. Yet detailed quantitative interpretations of the GRB brightness distribution have remained unresolved for several reasons. GRB rate-density evolution with cosmic time, and a range in GRB luminosities – inferred from redshift determinations but still not well sampled – complicate modeling of the relation (Krumholz, Thorsett, & Harrison 1998). Moreover it is necessary to consider in analyses newly identified untriggered bursts at the dim end of the distribution, where selection criteria (e.g., bandpass in rest frame) are different than for triggered bursts (Kommers et al. 1999; Stern et al. 1999). While these additional bursts may (or may not) support and extend the general flattening trend at very low peak fluxes, quantitative assessments of the log N – log PF relation become more difficult as subsets derived using different inclusion criteria are added. Thus the study of the GRB brightness distribution has begun to mature, following the usual course encountered for most astrophysical source groups, as the certain complexity of the phenomenon is unveiled, especially through definitive distance calibrations.

Two further related complications – also common in astronomical sources with ill-defined distance scales – arise in attempts to understand the log N – log PF relation: source subclasses and brightness bias. Subclasses of BATSE bursts have been identified which, while still exhibiting isotropic sky distributions, appear to have more homogeneous brightness distributions, closer to the expectation for sources uniformly populating Euclidean space. This may interpreted as evidence that the distributions as sampled represent relatively nearby bursters – possibly different populations of progenitors, or systematically underluminous GRBs. In fact, the only "HR diagram" for GRBs which has stood the test of time isolates subclasses by duration and gross spectral properties measured at BATSE energies (Kouveliotou, et al., 1996). Thus, short duration bursts, defined as having $T_{90} < 2$ s where $T_{90}$ is the time between the accumulation of 5% and 95% of the burst counts, have a more nearly –3/2 slope in their log N – log PF relation than do long bursts. The 2-s demarcation corresponds to bursts shortward of the valley in the bimodal burst duration distribution (Kouveliotou, et al. 1993).

Additionally, bursts selected according to the combined criteria of duration and BATSE hardness ratio, defined as the ratio of fluence in the 100–300 keV band to the 50–100 keV band (e.g. Pizzichini 1995; Kouveliotou 1996; Belli 1997) result in classes which exhibit different brightness distributions. In general, spectrally softer bursts (those with lower hardness ratios) exhibit a steeper, more nearly -3/2 slope in log N – log PF at lower peak fluxes. But, since short bursts form a definitive group in the GRB duration distribution, the question of subclasses in the brightness distribution would seem to be more cleanly investigated by considering only long bursts or short bursts and then separating bursts by spectral hardness. Recently, Tavani (1998) also concluded that long, hard bursts show the most significant deviation from a –3/2 slope. Tavani and the others argue that this is contrary to what is expected from the cosmological paradigm in which time dilation and redshift should conspire to produce the most significant flattening in the brightness distribution for the majority subclass of long, relatively soft bursts.

Pendleton et al. (1996) identified a class of bursts, similar to the spectrally soft (low hardness ratio) GRBs discussed above, characterized as those which exhibited a lack of emission above 300 keV. Pendleton et al. (1997, hereafter P97) substantially expanded this sample of NHE (no-high-energy emission) bursts, distinguished from HE (high-energy emission) bursts. P97 make a significant case that the brightness distribution of NHE bursts is more nearly homogeneous than that of HE bursts, suggesting that NHE burst sources are relatively nearby and systematically underluminous. P97 further identify NHE and HE emission peaks produced within the same GRB, and therefore suggest that the NHE bursts themselves are not a separate population of sources. Comparing the resulting brightness distributions for NHE and HE bursts, they estimate the ratio of their spatial densities and indicate that the NHE spectral component of GRBs may simply be less luminous but more broadly beamed than the HE emission.

This work presents a different interpretation of the relative brightness distributions of HE and NHE bursts. In general, we find that for a cosmological population of bursters, NHE bursts and other classes of long, soft BATSE GRBs would be expected to show more nearly homogeneous brightness distributions when brightness biases are considered. We therefore suggest that these brightness distributions are not evidence for a distinct and nearby population of bursters nor do they represent evidence for differential beaming that would contribute to the diverse spectral properties of prompt GRB emission (Meszaros, Rees, & Wijers 1998). In section 2 we describe our sample of bursts selected from the existing archive of BATSE data. In section 3 we determine their peak fluxes and find the corresponding NHE and HE brightness distributions. In section 4 we discuss the synthesis of a spectrally diverse cosmological population of bursts based on observed bright BATSE GRBs and accounting for appropriate signal-to-noise (s/n) reductions and redshift. The synthetic NHE brightness distributions are compared to those derived in section 3. For the synthetic population, a GRB peak flux versus redshift relation based on Bonnell et al. (1997) is adopted which effectively reproduces the observed trend of spectral hardness versus peak flux (Mallozzi et al 1995). In section 5 we discuss our results in terms of their implications for GRB studies and from the general point of view of the problem of brightness bias in astronomy.

## 2. THE GRB SAMPLE

Our goal is to determine how closely the brightness distribution of soft GRBs, exemplified by an apparent NHE class, can be reproduced by: (1) adjusting the s/n levels of bright bursts which do exhibit significant emission > 300 keV appropriate to decreasing peak flux; (2) redshifting the burst spectral energy distributions according to peak flux versus time dilation measurements, and (3) calculating what fraction of these synthetic bursts have no remaining statistically significant HE emission. Since we wish to address the problem in the most self-consistent manner, we choose not to include short bursts, with $T_{90} < 2$ s, in the sample for several reasons. Primarily, the sensitive tests used here for HE emission examine the background-subtracted count rate $> 300$ keV at 1-s resolution in the $T_{90}$ interval. Yet, the pulses in short bursts can often be very narrow, spanning a much smaller fraction of a 1-s sample than pulses in long bursts (Norris 1995), and so a large bias against detecting HE emission would result for short bursts. Moreover, the relation we use to effect spectral redshift is derived from time-dilation measurements of long bursts, and may not be applicable to short bursts.

Short bursts alone may indeed constitute a class separate from cosmological long bursts. In addition to constituting a separable short mode in the burst duration distribution, short bursts have systematically harder spectra than long bursts. As discussed above, their brightness distribution – which more nearly follows a -3/2 power-law – suggests that a closer population of source objects is sampled than for long bursts. To date, there are no counterpart detections or conventional redshift measurements for short bursts, and thus no definitive evidence that they lie at cosmological distances.

Hence we elected to study the NHE question using only long bursts ($T_{90} > 2$ s). Through May 15, 1997 there were 1821 BATSE bursts available in the Compton GRO Science Support Center archive (up to trigger 6230). Following procedures described in Bonnell et al. (1997), we constructed concatenated time profiles for all bursts where possible, requiring timing overlap for the 1.024-s continuous DISCLA data and the 0.64-ms PREB and DISCSC data, which are recorded by the Large Area Detectors (LADs). We then fitted and subtracted backgrounds for each of the four LAD energy channels (25–50, 50–100, 100–300, $> 300$ keV), and estimated $T_{90}$ durations using a standard noise equalization and thresholding procedure, also described in Bonnell et al., for all bursts with sufficient data preceding and following the burst. The preparation procedure – concatenation, background fitting and subtraction, and duration measurement – yielded a usable sample of 1010 long bursts.

## 3. PEAK FLUX AND NHE DETERMINATIONS

GRB peak fluxes in the BATSE catalog are computed on the three trigger timescales (64 ms, 256 ms, and 1024 ms) in the 50–300 keV energy range (LAD channels 2+3) following an algorithm described in published BATSE catalogs (see Fishman et al. 1994) and detailed, for example, by Pendleton et al. (1996). A direct inversion of the detector response matrices is used to convert the four-channel background-subtracted LAD counts to photons cm$^{-2}$ s$^{-1}$.

Peak fluxes determined on the 256-ms timescale have been shown to be adequate representations of the instantaneous peak in long bursts (Bonnell et al. 1997). We elected to measure the burst peak flux (PF) on a 256-ms timescale for all bursts in this sample using our own background determinations and an independently coded algorithm designed to follow the published method used for the BATSE catalog. Figure 1 shows the ratio of our PF to the PF tabulated in the online BATSE catalog versus trigger number.

The agreement between our PF values and the BATSE catalog values is reasonable; Figure 1 reveals only a small (~< 10%) systematic difference. Note, however, the larger dispersion up to trigger 1466 which corresponds to the end of the BATSE 1B catalog. The BATSE catalog PF values for these earlier bursts were calculated using an obsolete algorithm (Pendleton private comm.) and have not been updated. Since all our PF values are calculated using a single approach based on the current BATSE algorithm, and since we use the same background-subtraction for the PF and NHE computations, we have adopted our own PF values for constructing GRB brightness distributions. The resulting cumulative brightness distribution, log N – log PF, for the 1010 long bursts is plotted in Figure 2.

For each burst in the sample we then determined if significant HE emission is present within the $T_{90}$ interval. The background-subtracted time profile (signal) of each long burst at its original s/n level was binned to 1.024-s resolution and two criteria for identifying HE emission above a chosen threshold were applied. If either one or both were satisfied, the burst was declared HE; otherwise it was declared NHE. The HE criteria were:

(1) the signal summed over the entire $T_{90}$ interval was $> 4\,\sigma$ above the total variance in the same interval;
(2) the signal summed over the subinterval (within the $T_{90}$ interval) with the highest contiguous positive residuals (rms) was $> 4\,\sigma$ above total variance in the same subinterval.

The total variance included the signal and background variances in quadrature. The second criterion was designed to isolate the most intense pulse structure and thereby exclude low-intensity or near-background level samples, which tend to diminish the significance of intervals with HE emission. Almost invariably, the second criterion was more sensitive to HE emission. However, infrequently only the first criterion was satisfied, indicating long stretches at low intensity which summed to significant emission. A third criterion – a single 1-s sample $4\,\sigma$ above background – never signaled HE emission when both criteria (1) and (2) failed; consequently we did not utilize this less sensitive measure. A similar NHE/HE discrimination procedure is described in Norris et al. (1998).

The remaining two histograms plotted in Figure 2 are the NHE and HE groups, comprising 571 and 439 bursts, respectively, as identified by our two criteria. The NHE curve commences below a PF of ~ 10 photons cm$^{-2}$ s$^{-1}$; there are no brighter bursts which are categorized as NHE. The steepness of the NHE brightness distribution – *apparently* counterintuitive given that the curve for the whole set of long bursts becomes flatter on the dim end – is the basis for claims of a distinct group of dim, soft bursts with a more homogeneous brightness distribution. However, as we proceed to demonstrate, it is straightforward to construct a very similar synthetic NHE curve using bright bursts with HE emission that have had their s/n levels decreased and their spectra redshifted.

## 3. SYNTHESIS OF THE NHE BRIGHTNESS DISTRIBUTION

To synthesize an extensive brightness distribution of GRBs we constructed a procedure to dim and redshift a set of bright bursts. The bright burst set comprised only the brightest, long bursts which have 16-channel MER and CONT data covering the whole burst. Since the intent is to adopt empirical burst photon spectra and redshift them, the moderate energy resolution of the 16-channel data types was needed to adequately characterize the shape of the spectrum and extend it to relatively high energies. As might be expected, the BATSE 4-channel data, with only 2 channels above 100 keV, proved to be inadequate to determine burst spectral characteristics for the ranges critical in this synthesis procedure. Above a PF of ~ 8.0 photons cm$^{-2}$ s$^{-1}$ – sufficiently bright that virtually all long bursts have HE emission – we identified 46 bursts up to trigger 5490 with usable 16-channel data. These burst's BATSE trigger numbers and PF determinations are given in Table 1.

We then determined at what decreased PF level the HE emission becomes undetectable in the corresponding 4-channel data, according to the same HE criteria used for the whole set of long bursts. To accomplish this step, the s/n level was reduced in the 4-channel time profiles of the bright bursts as follows. For each of 40 logarithmic bins spanning 4 decades (0.1 – 10$^3$ photons cm$^{-2}$ s$^{-1}$) in the log N – log PF, we reduced the signal time profile, according to Norris et al. (1994), adding simulated Poisson noise per DISCSC channel to render the s/n level equal to that of a randomly chosen burst in our observed sample within a given PF bin. This signal reduction step produced 1840 synthetic time profiles (40 PF steps × 46 bursts). The signal level in channel 4 was further reduced to account for the spectral redshift appropriate to the PF bin. This redshifting procedure is described in detail below.

We computed a grid of redshifted spectra for each burst, spanning 1 < (1+z) < 5 in steps of Δz = 0.2. We deconvolved the 16-channel energy-loss spectrum using the appropriate detector response matrices to produce a photon spectrum under two different assumptions, the Band (1993) model and a broken power-law. For each step in the z grid, we redshifted the two model-dependent photon spectra, and redetected with 4-channel detector response matrices. As a function of redshift step in the grid, we then computed the count ratios, $f_i(z)$, in the 4 channels – after : before the redshift procedure. To determine which $f_i(z)$ to use for each of the 40 PF steps, we used the correspondence in Table 2 between our measurement of redshift-corrected time-dilation factor (TDF) and the PF lower bound. For Table 2 we used TDFs measured for the present sample of long bursts; each PF group contains ~ 140 bursts. For the PF groups 0.32–0.68 and 0.68–0.95, the TDF analysis includes a correction for BATSE's diminishing sensitivity to slowly rising bursts. The correction is estimated by computing TDFs only for the decaying portion of the average, peak-aligned time profiles (Norris, Bonnell, & Watanabe, 1999b). We then chose for each PF step and bright burst the $f_i$ to be utilized from the grid, based on the redshift factor read out from Table 2. Finally, the $f_i$ were used to adjust the signal level in channel 4 produced by the s/n-reduction procedure, thereby simulating the effect of redshift:

$$C_4(t)_{redshifted} = C_4(t) \times (P_2 + P_3)/(f_2 P_2 + f_3 P_3) \times f_4 \qquad (1)$$

where $C_4(t)$ is the time profile in channel 4 produced in the s/n-reduction procedure, and the $P_i$ are peak count rates determined on a 1-s timescale at the original s/n levels of the 46 bright bursts. The factor $(P_2 + P_3)/(f_2P_2 + f_3P_3)$ renormalizes the time profile in channel 4, based on the ratio of peak counts in the sum of channels 2 and 3, before : after redshift. This renormalization factor is necessary since the BATSE experiment trigger is based on the sum of counts in channels 2 and 3. Thus the effect of the redshift procedure is to redistribute counts in the four channels, with the sum of counts in the peak-flux energy window unchanged. The reduction factor for channel 4, $f_4$, is always less than unity.

We note that the final results are not very sensitive to the specific parameterization employed to represent the observed correlation of burst spectra hardness with peak flux, as reported by Mallozzi et al. 1995). To synthesize the NHE brightness distribution, it is merely necessary to adopt some parametric relationship between PF and spectral hardness that adequately approximates the observed dependence. Since our time-dilation factors are in reasonable agreement with the Mallozzi et al. findings, and since our results were computed for only long bursts, we utilized our TDFs.

Figure 3 illustrates the resulting synthetic cumulative brightness distributions computed for the two model spectra assumed. The curve for the Band (power-law) model lies higher (lower). The observed NHE brightness distribution is also plotted with a thicker line. The same criteria were applied to the determination of NHE for the synthetic bursts and the observed sample. The two synthetic curves bracket the actual NHE curve. Figure 4 plots the same brightness distribution data as a histogram of number of bursts in each peak flux bin (the differential distribution).

## 4. DISCUSSION

The synthetic distributions computed for the two model spectra are seen in Figure 3 to have a general slope that falls below a -3/2 power law at significantly lower peak fluxes than the distribution for all long bursts or even all long HE bursts. This result is clearly qualitatively similar to reports in previous works for the putative NHE class (P97). Thus we have demonstrated that the appearance of the observed NHE brightness distribution can be understood as a natural consequence of brightness bias: decreasing s/n level and softer spectra, presumably redshifted, combine to increase the apparent fraction of NHE bursts at low peak fluxes. The factors considered in our NHE burst synthesis derive from a single cosmological distance scale for bursts and from the bright bursts themselves. It was not necessary to assume a systematically underluminous population of bursts occupying nearby Euclidean space. Instead, the bright bursts utilized to make synthetic bursts manifest a range in spectral diversity, such that when they are dimmed to successively lower peak fluxes and redshifted, we find the rate of occurrence of synthetic bursts with NHE to be very similar to the observed relation. The PF vs. redshift algorithm we employed is an empirical one which parameterizes the observed properties of long bursts, independently of a specific cosmological model. The cause of spectral softening need not be cosmological redshift for our spectral softening parameterization to be effective; rather the algorithm merely needs to reflect the general softening trend with peak flux. In fact, the time-dilation trend is similar to the apparent spectral redshift trend reported by Mallozzi et al. (1995).

A detailed comparison between the observed and synthetic NHE curves illustrated in Figure 3 and Figure 4 indicates that the high PF ends of the brightness distributions of the synthetic burst curves contain only slightly fewer NHE bursts than the observed distribution. While this mismatch is insensitive to the spectral models adopted – and to the threshold used to realize the NHE criterion – it could be adjusted, for example, by choosing a different initial set of bright bursts which define the spectral diversity used in the burst synthesis. We chose not to tailor the burst synthesis in this manner since the differential distributions in Figure 4 emphasize that the residual differences involve small numbers of bursts.

At the dim end of the PF distribution, the synthesis involving Band model spectra produced too many faint NHE bursts while the power-law spectral model produced too few. Combined, the two models bracket the observed distribution. Both functions are empirical fits and can provide spectral curvature in the energy band of BATSE observations and peak flux determinations. Thus the true burst spectrum may lie between a smoothly curving Band model and the similar but sharply broken power-law model. More likely, an additional selection effect could be operating that we have not taken into account.

P97 describe a sample with another selection criterion – intervals with NHE emission in bright bursts (which also have intervals of HE emission) – for which the associated peak-flux measures exhibit a locus similar to a -3/2 power-law in their brightness distribution. This appearance can be understood in much the same way as the brightness distribution for whole bursts which have NHE emission throughout, that is, as brightness bias. Figure 3 also illustrates the log N – log PF with the same procedures applied as described in section 3, with and without the redshift step. Merely decreasing the s/n levels yields ~ 2/3 of the number of observed NHE bursts, but with virtually the same slope in the brightness distribution as when the redshift step is included. Thus it appears that the second NHE phenomenon described in P97 could arise merely from considering the dimmer portions of bright bursts with HE emission.

It has been remarked that the relatively steep slope of the NHE brightness distribution (and other similarly defined distributions which require soft spectra) is counterintuitive, and difficult to understand from first principles. Since the general GRB brightness distribution flattens at lower peak fluxes, and since softer bursts only begin to prevail in the same regime, how can the NHE brightness distribution be so steep? The clear answer lies in the definition of log N – log PF for NHE bursts (and spectrally similar bursts): (1) the cumulative nature of N(>PF) necessarily yields a decreasing function, with a lower PF threshold for inclusion (due to instrumental and extrinsic factors) and it so happens that the lower end of the PF range is inhabited by softer bursts; while (2) the additional requirement of NHE imposes an upper PF threshold for inclusion! The apparent NHE class is therefore trapped between two thresholds since the ordinate being plotted is not the usual N(>PF), but rather

$$N( > PF[50\text{–}300 \text{ keV}] \text{ .AND. } F[>300 \text{ keV}] < n\sigma ) \qquad (2)$$

where $F[>300 \text{ keV}] < n\sigma$ can be replaced by any similar requirement that results in a soft spectrum. The combined requirements – that a burst have *some* measurable flux in the energy range 50–300 keV, but *essentially none* in the adjacent higher range – necessarily gives rise to the fast accumulation of NHE bursts, and their rapid exhaustion, in a narrow PF range. Indeed, low

PF is already the regime of soft-spectrum bursts, by virtue of which the NHE PF range is restricted once. The NHE PF range is restricted a second time by absence of HE (dimness), moving the upper threshold towards the lower threshold. Thus, only a relatively small fraction of bursts with similar PF values will satisfy both requirements, and their accumulation as PF decreases will occur rapidly. It would be surprising if such a constrained brightness distribution resulted in an exact –3/2 relation at all. In fact, the synthetic and observed NHE curves in Figure 3 are seen to be slightly steeper than a -3/2 power law. The nearness to a –3/2 slope is then just a coincidence analogous to the fortuitously long range of the –3/2 region for bright bursts, which is now ascribed to the convolution of geometry and evolution of the cosmic star formation rate.

## 5. CONCLUSIONS

We find that the putative NHE class, and similar groups of bursts defined by soft spectra, have steep brightness distributions which can be understood in terms of brightness bias. In particular, synthetic bursts, generated by signal-to-noise equalization and spectral reddening of bright bursts to approximate characteristics of dim bursts exhibit a distribution statistically very similar to that observed for the NHE class. Consequently, we have demonstrated that an NHE-like brightness distribution can be *naturally expected* to commence with a steep slope, and at significantly lower fluxes than long, hard (HE) bursts. This steep slope is due to brightness bias at lower peak fluxes and associated higher relative redshifts (or empirical reddening).

Interpretations of soft burst brightness distributions as evidence for distinct subclasses or as a reflection of intrinsic source characteristics of gamma-ray bursters are therefore seriously challenged. Their nearly –3/2 slope brightness distributions can not be considered as clear evidence for sub-luminous or nearby populations of sources, or as evidence of differential beaming as has been suggested for the NHE class. The similarly steep brightness distribution of NHE emission within HE bursts reported by P97 can be understood in terms of brightness bias as well. We also note that brightness bias makes the NHE character of the emission in GRB 980425 (BATSE trigger 6707) difficult to regarded as a defining characteristic of a GRB-SN type class (Norris, Bonnell, & Watanabe 1999).

Further claims for detailed interpretations of GRB brightness distributions based on gross spectral characteristics must account for this demonstrably large brightness bias at low peak fluxes.

## ACKNOWLEDGMENTS

The background fits and concatenated 64-ms data used in this analysis can be found at these web sites, maintained by the Compton GRO Science Support Center:

http://cossc.gsfc.nasa.gov/cossc/batse/batseburst/sixtyfour_ms/bckgnd_fits.htm
http://cossc.gsfc.nasa.gov/cossc/batse/batseburst/sixtyfour_ms/

Table 1. Peak Fluxes for Bright BATSE GRBs

| Trigger Number | Burst | Peak Flux (error) 50 – 300 keV | | Peak Flux (error) > 300 keV | |
|---|---|---|---|---|---|
| 105 | GRB 910421 | 12.86 | (0.32) | 0.53 | (0.18) |
| 143 | GRB 910503 | 62.18 | (0.58) | 22.91 | (0.51) |
| 219 | GRB 910522 | 19.41 | (0.34 | 3.86 | (0.26) |
| 249 | GRB 910601 | 38.08 | (0.41) | 12.47 | (0.36) |
| 451 | GRB 910627 | 10.20 | (0.32) | 0.69 | (0.22) |
| 543 | GRB 910717 | 9.88 | (0.26) | 0.96 | (0.20) |
| 678 | GRB 910814 | 10.24 | (0.27) | 11.88 | (0.45) |
| 999 | GRB 911104 | 13.09 | (0.32) | 4.49 | (0.32) |
| 1025 | GRB 911109 | 16.68 | (0.40) | 2.03 | (0.26) |
| 1085 | GRB 911118 | 28.82 | (0.44) | 1.35 | (0.19) |
| 1122 | GRB 911127 | 11.54 | (0.27) | 1.52 | (0.18) |
| 1141 | GRB 911202 | 10.62 | (0.29) | 2.56 | (0.27) |
| 1157 | GRB 911209 | 10.39 | (0.28) | 1.99 | (0.24) |
| 1425 | GRB 920221 | 8.57 | (0.26) | 1.07 | (0.19) |
| 1440 | GRB 920226 | 13.14 | (0.30) | 2.10 | (0.25) |
| 1709 | GRB 920718 | 12.00 | (0.29) | 0.74 | (0.16) |
| 1886 | GRB 920902 | 15.91 | (0.33) | 7.23 | (0.40) |
| 2067 | GRB 921123 | 17.90 | (0.37) | 2.96 | (0.27) |
| 2090 | GRB 921209 | 9.15 | (0.25) | 1.40 | (0.20) |
| 2156 | GRB 930201 | 16.91 | (0.34) | 3.60 | (0.30) |
| 2329 | GRB 930506 | 40.79 | (0.52) | 10.63 | (0.49) |
| 2431 | GRB 930706 | 33.85 | (0.43) | 4.03 | (0.31) |
| 2533 | GRB 930916 | 8.94 | (0.27) | 3.82 | (0.30) |
| 2537 | GRB 930922 | 25.42 | (0.41) | 2.20 | (0.24) |
| 2586 | GRB 931014 | 8.05 | (0.23) | 0.71 | (0.19) |
| 2611 | GRB 931031 | 29.96 | (0.45) | 9.85 | (0.45) |
| 2797 | GRB 940203 | 8.09 | (0.25) | 1.84 | (0.25) |
| 3057 | GRB 940703 | 31.97 | (0.45) | 14.55 | (0.54) |
| 3115 | GRB 940810 | 10.71 | (0.30) | 2.92 | (0.31) |
| 3138 | GRB 940826 | 15.97 | (0.36) | 2.46 | (0.28) |
| 3227 | GRB 941008 | 15.81 | (0.34) | 5.22 | (0.34) |
| 3241 | GRB 941014 | 12.18 | (0.29) | 1.13 | (0.18) |
| 3245 | GRB 941017 | 12.38 | (0.30) | 3.90 | (0.31) |
| 3253 | GRB 941020 | 14.03 | (0.31) | 2.66 | (0.26) |
| 3290 | GRB 941121 | 9.86 | (0.26) | 0.54 | (0.18) |
| 3415 | GRB 950211 | 8.52 | (0.28) | 0.83 | (0.18) |
| 3458 | GRB 950305 | 8.31 | (0.26) | 2.51 | (0.260 |
| 3481 | GRB 950325 | 21.02 | (0.39) | 5.90 | (0.37) |
| 3488 | GRB 950401 | 8.62 | (0.28) | 2.11 | (0.27) |
| 3491 | GRB 950403 | 32.97 | (0.54) | 3.43 | (0.29) |
| 3658 | GRB 950701 | 12.12 | (0.29) | 3.48 | (0.31) |
| 3870 | GRB 951016 | 14.06 | (0.35) | 2.27 | (0.32) |
| 3891 | GRB 951102 | 12.75 | (0.31) | 3.59 | (0.33) |
| 3954 | GRB 951213 | 9.71 | (0.34) | 2.55 | (0.34) |
| 5486 | GRB 960605 | 9.03 | (0.27) | 1.80 | (0.26) |
| 5489 | GRB 960607 | 9.22 | (0.27) | 1.20 | (0.20) |

Peak fluxes determined on a 256 ms timescale are given in photons·cm$^{-2}$·s$^{-1}$ for the indicated energy range. Corresponding errors are in parenthesis.

Table 2. Peak Flux and Time-Dilation Correspondence

| | | | | | | |
|---|---|---|---|---|---|---|
| Peak Flux (lower bound)[1] | 4.6 | 2.25 | 1.37 | 0.95 | 0.68 | 0.32 |
| Time-Dilation Factor | 1.00 | 1.50 | 1.80 | 2.00 | 2.20 | 2.75 |

[1] In photons cm$^{-2}$ s$^{-1}$ (50 – 300 keV) on 256-ms timescale

Figure Captions

Figure 1: The ratio of our PF (peak flux) to the PF tabulated in the online BATSE catalog versus trigger number. Horizontal lines are drawn at PF ratios of 1.0 and 0.9. Note the larger dispersion prior to trigger 1466 (indicated by the vertical line) which corresponds to the end of the BATSE 1B catalog. After trigger 1466, most PF ratios lie between 1.0 and 0.9 (good agreement) and show no clear trend.

Figure 2: The resulting cumulative brightness distributions, log N – log P, for the 1010 long bursts analyzed here. The thick solid curve represents all bursts, the thin solid curve only NHE bursts, and the dotted curve only HE bursts. The straight line indicates a reference slope of -3/2. The NHE burst distribution is steeper, closer to a -3/2 slope at lower peak fluxes, than the HE or total burst distributions.

Figure 3: The synthetic NHE brightness distributions compared to the observed NHE distribution (thick solid curve). The thin solid curve is the synthetic distribution produced by adding only noise to bright bursts. The other curves add noise *and also* redshift the bright bursts according to two adopted photon spectrum functions; a Band function (dotted) and a broken power-law function (dashed). The curve for the Band (broken power-law) function lies higher (lower) than the observed distribution at low peak fluxes. Note that all these distributions are slightly steeper than the straight reference line with -3/2 slope.

Figure 4: The differential NHE brightness distributions compared (synthetic vs. observed). The solid curve represents the observed distribution. The synthetic distributions are the dashed (broken power-law) and dotted (Band function) curves as described for Figure 3.

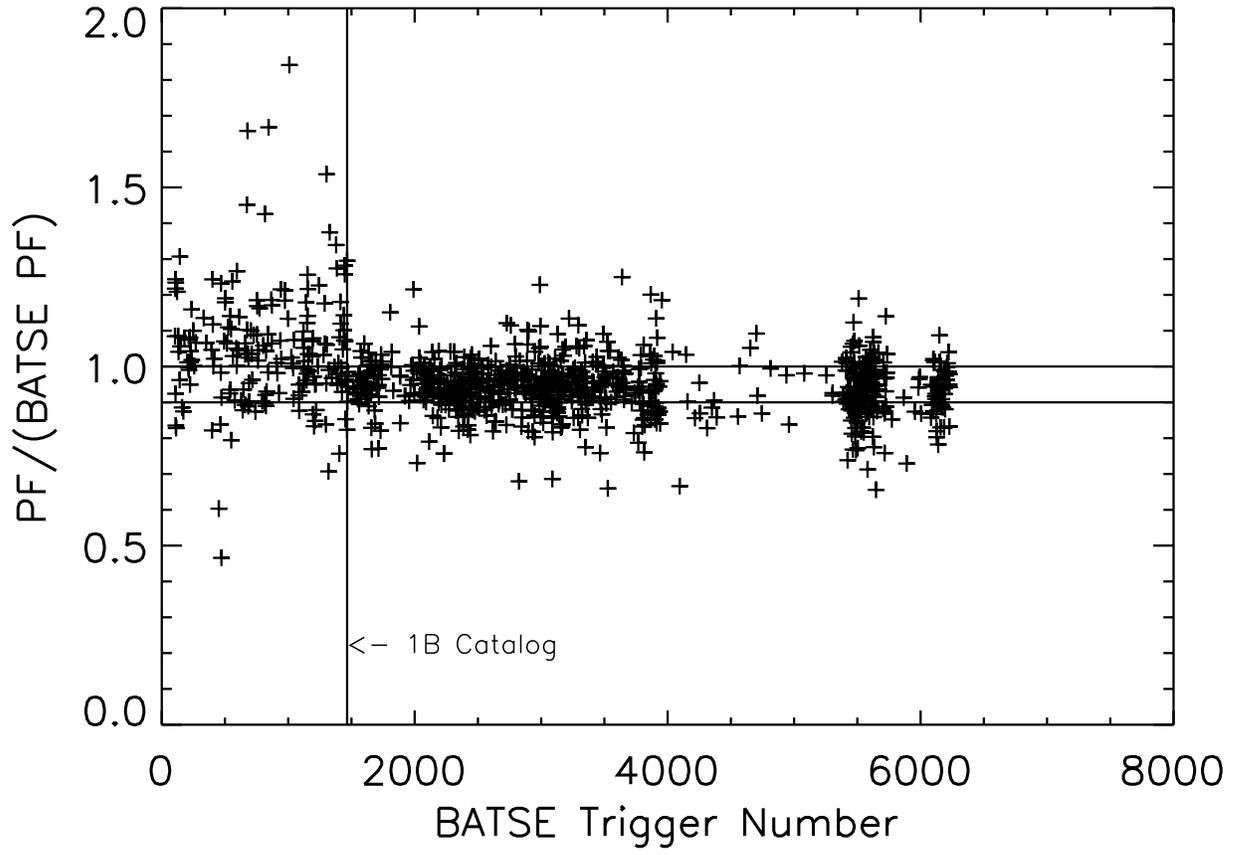

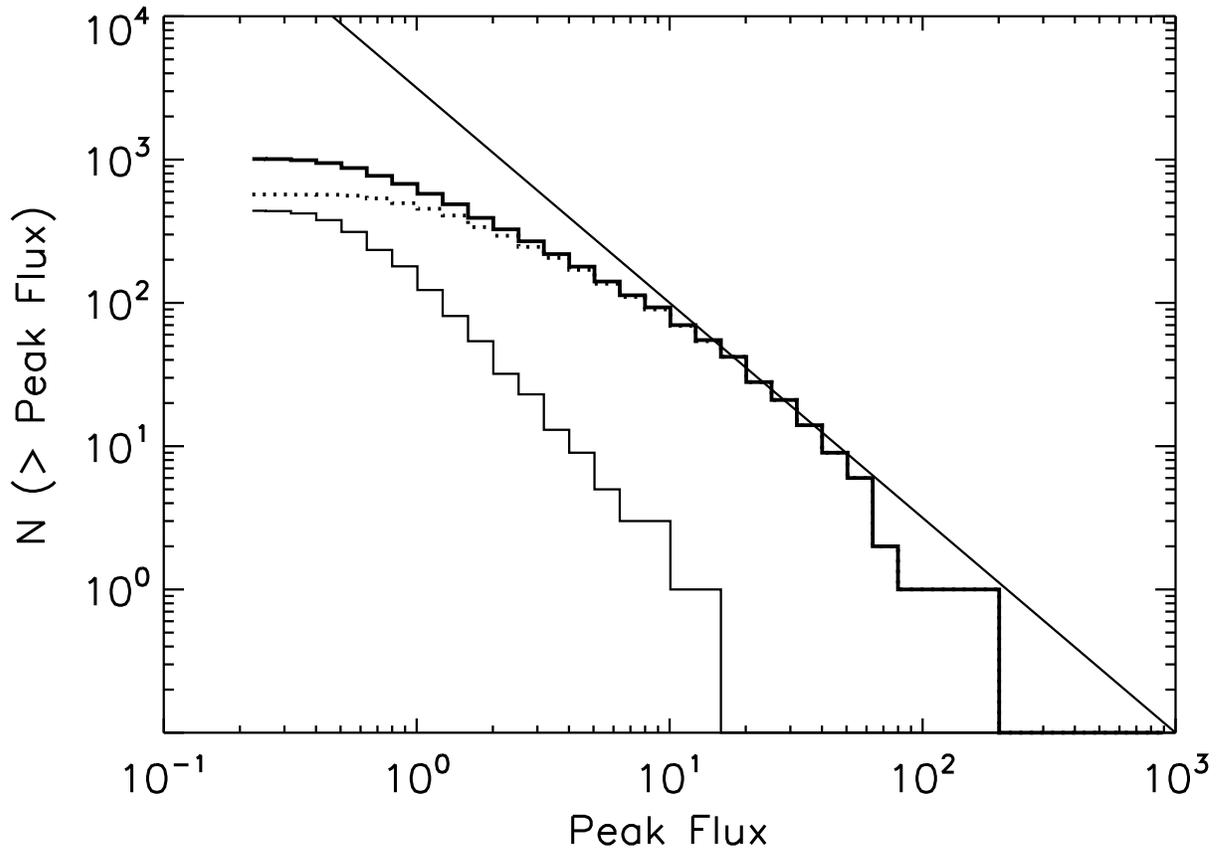

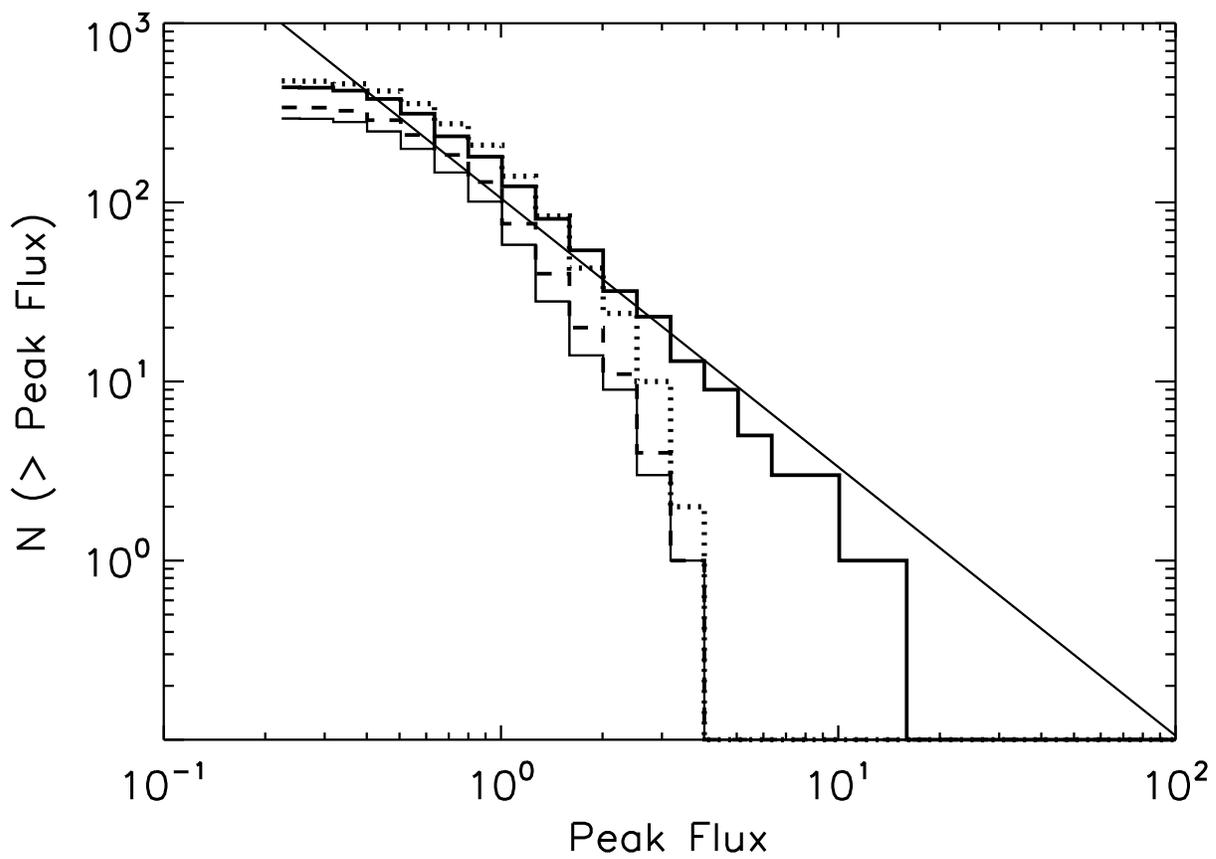

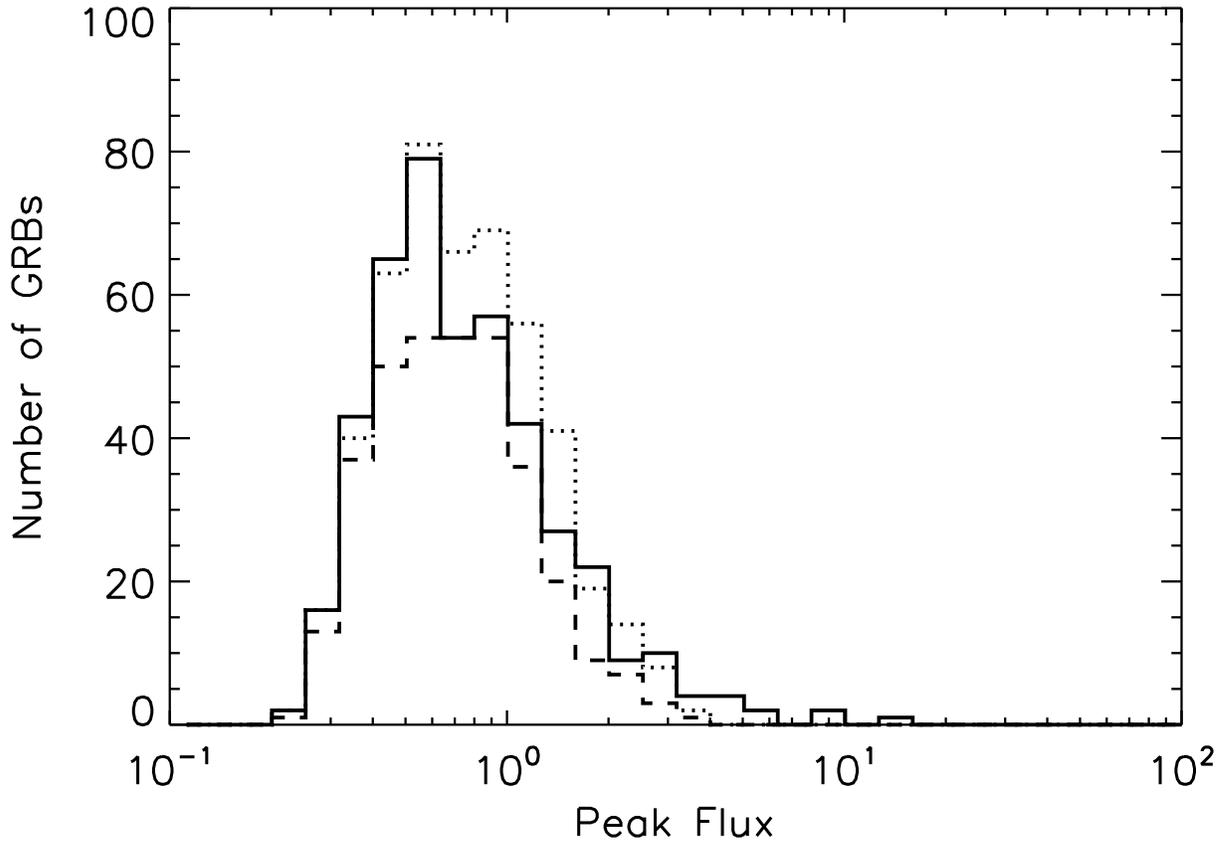